# The Optimal Spatially-Smoothed Source Patterns for the Pseudospectral Time-Domain Method

Zhili Lin

***Abstract*—** **Spatially-smoothed sources are often utilized in the pseudospectral time-domain (PSTD) method to suppress the associated aliasing errors to levels as low as possible. In this work, the explicit conditions of the optimal source patterns for these spanning sources are presented based on the fact that the aliasing errors are mainly attributed to the high spatial-frequency parts of the time-stepped source items and subsequently demonstrated to be exactly corresponding to the normalized rows of Pascal's triangle. The outstanding performance of these optimal sources is verified by the practical 1-D, 2-D and 3-D PSTD simulations and compared with that of non-optimal sources.**

## I. INTRODUCTION

THE pseudospectral time-domain (PSTD) method has been widely applied to simulate various electromagnetic and acoustic problems since its emergence at the end of last century [1], [2]. Basically, it uses discrete Fourier transform (DFT) or Chebyshev transforms to calculate the spatial derivatives of the electric and magnetic field components both arranged in the same positions in an unstaggered space lattice of unit cells. Because the spatial-differencing process in PSTD method converges with infinite order accuracy for a low sampling density of two points per shortest wavelength, it renders lower numerical phase-velocity errors than those of the finite-difference time-domain (FDTD) method and therefore allows problems of much larger electrical size to be modeled [3].

However, the DFT, implemented by the famous fast Fourier transform (FFT), has difficulty in correctly representing the Kronecker delta function. With a single-cell source applied, zigzag wiggles are arising apparently on the excited source waves, referred to as the Gibbs phenomenon [4]. To alleviate these aliasing errors, spatially smoothed sources spanning a few (4~6) cells in each coordinate direction were proposed by Liu [5], but without further details on their optimal patterns. The compact two-identical- cell source has also been investigated by Lee and Hagness [6]. However, their study is only based on the method of comparison and the specific guidelines for the optimal source patterns are still unreported.

In this work, the explicit conditions for the optimal source patterns with lowest levels of aliasing errors are investigated according to the fact that the aliasing errors are mainly attributed to the high spatial-frequency components of the added source items at each time step. We further demonstrate that the amplitude distributions of these optimal source patterns are exactly corresponding to the normalized rows of Pascal's triangle. The practical 1-D, 2-D and 3-D PSTD simulations are conducted to verify the validity and performance of these proposed sources.

## II. FORMULATION

Supposing a TEM plane wave propagating in the free space along the $x$ axis with the electric field vector oriented in the $z$ direction. The 1-D grid space is composed of $N$ cells indexed by $i = 0, 1, 2, ..., N-1$. According to the PSTD algorithm, updating equations for the *normalized* electric and magnetic fields excited by the source item $S^{n+1/2}$ are given by

$$E_z^{n+1/2} = E_z^n + c_x \boldsymbol{F}^{-1}\{K_x \cdot \boldsymbol{F}[H_y^{n+1/2}]\}, \quad (1)$$

$$E_z^{n+1} = E_z^{n+1/2} + S^{n+1/2}, \quad (2)$$

$$H_y^{n+3/2} = H_y^{n+1/2} + c_x \boldsymbol{F}^{-1}\{K_x \cdot \boldsymbol{F}[E_z^{n+1}]\}, \quad (3)$$

where $c_x = c\Delta t / \Delta x$ with $c$, $\Delta t$ and $\Delta x$ being the speed of light in vacuum, the step time and cell size used in a specific problem, respectively; $\boldsymbol{F}$ and $\boldsymbol{F}^{-1}$ are the forward and inverse DFT defined by

$$X(k) = \sum_{i=0}^{N-1} x(i) e^{-j\frac{2\pi}{N}ik}, \quad k = 0, ..., N-1 \quad (4)$$

$$x(i) = \frac{1}{N}\sum_{k=0}^{N-1} X(k) e^{j\frac{2\pi}{N}ki}, \quad i = 0, ..., N-1 \quad (5)$$

Manuscript received April 15, 2008. This work was supported by the program between the China Scholarship Council and the Royal Institute of Technology, Sweden.
Z. Lin is currently with the School of Instrument Science and Opto-electronics Engineering, Beihang University, Beijing, 100191, China (e-mail: zllin2008@gmail.com ).

and the differential factor $K_x$ is given by

$$K_x(k) = \begin{cases} j2\pi k/N, & k = 0,...,N/2-1 \\ 0, & k = N/2 \\ j2\pi(k-N)/N. & k = N/2+1,...,N-1 \end{cases} \quad (6)$$

Assume that the superimposed item in (2) is

$$S^{n+1/2} = [0,...,0,a_0,a_1,...,a_{m-1},0,...,0]s^{n+1/2}, \quad (7)$$

comprising $m$ source cells, say, the $i_0, i_1, ..., i_{m-1}$ th cells with normalized amplitudes satisfying

$$a_0 + a_1 + ... + a_{m-1} = 1. \quad (8)$$

In fact, both the soft and hard source cases have been considered in (2). That is, (7) is standing for a soft source $S_s^{n+1/2}$ by letting $s^{n+1/2} = f^{n+1/2}$ or standing for a hard source $S_h^{n+1/2}$ by letting

$$s^{n+1/2} = f^{n+1/2} - \sum_{k=0}^{m-1} E_z^{n+1/2}(i_k), \quad (9)$$

where $f^{n+1/2}$ denotes the temporal driving function $f(t)$ at time step $n+1/2$.

The pattern of the source item $S^{n+1/2}$ in (2) is crucial to the problem we concerned. With (7), the spatial spectrum $S(k)$ of the pattern $[a_0, a_1, ... a_{m-1}]$ is

$$S(k) = F[S^{n+1/2}/s^{n+1/2}] = \sum_{l=0}^{m-1} a_l e^{-j\frac{2\pi}{N}ki_l}, \quad (10)$$

where $k$ is the discrete spatial frequency and physically equivalent to $1/(2\Delta x)$ for $k = N/2$. Based on (1-3), we can further define that the dominant aliasing errors of the magnetic and electric field independently caused by $S^{n+1/2}$ are

$$\sigma_H^{n+3/2} = c_x s^{n+1/2} F^{-1}\{K_x \cdot S\}, \quad (11)$$

$$\sigma_E^{n+2} = c_x^2 s^{n+1/2} F^{-1}\{K_x^2 \cdot S\}, \quad (12)$$

on the cells away from the source region. From (11) and (12), we note that the spatial spectrums of the aliasing errors $\sigma_H^{n+3/2}$ and $\sigma_E^{n+2}$ are proportional to the products of $K_x$ and $S$, $K_x^2$ and $S$, respectively. In Fig.1, we show the magnitude of the spatial spectrums of $K_x$ and $K_x^2$ with $N = 128$, both possessing ample high-spatial-frequency components. Thus, to suppress the aliasing errors, the pattern $[a_0, a_1, ... a_{m-1}]$ should be optimally designed to make $S(k)$ being a decreasing discrete function, as rapidly as possible, in the region $0 \le k \le N/2$ and become zero at $k = N/2$. This requires $S(N/2) = 0$ and $|S|$ should simultaneously reach their minimum values in the high discrete spatial-frequency range near $N/2$, such as at $N/2-1$, $N/2-2$ and so on. By the Lagrange multiplier method, we have the following necessary conditions for the optimal source patterns

$$S(N/2) = \sum_{l=0}^{m-1} (-1)^l a_l = 0, \quad (13)$$

$$\frac{\partial L_{N/2-p}}{\partial a_0} = \frac{\partial L_{N/2-p}}{\partial a_1} = ... = \frac{\partial L_{N/2-p}}{\partial a_{m-1}} = 0 \quad (14)$$

for $p = 1, 2, ..., p_{\max}$, where the Lagrange function $L_{N/2-p}$ is defined by

$$L_{N/2-p}(a_0,...,a_{m-1},\lambda_p) = \left|S(\frac{N}{2}-p)\right| + \lambda_p S(\frac{N}{2})$$
$$= \left|\sum_{l=0}^{m-1} a_l e^{-j\frac{2\pi}{N}(\frac{N}{2}-p)l}\right| + \lambda_p \sum_{l=0}^{m-1} (-1)^l a_l \quad (15)$$

with $\lambda_p$ being the Lagrange multiplier. From (15), the $m$ equations in (14) for each $p$ can be simplified as

$$\sum_{q=0}^{m-1} a_q \cos[\frac{2\pi}{N}(q-l)(\frac{N}{2}-p)] = (-1)^{l+1} \lambda_p \left|S(\frac{N}{2}-p)\right| \quad (16)$$

for $l = 0, 1, ... m-1$. We note that the rank of the coefficient matrix of the $m$ equations in (16) is 2, so only two among them are independent. Thus to determine a specific source pattern, the limit of $p$ is up to $p_{\max} = (m-2)/2$ for an even $m$ or $p_{\max} = (m-1)/2$ for an odd $m$, which guarantees the magnitudes of $S$ at the discrete spatial frequencies, $N/2, N/2-1, ..., N/2-p_{\max}$ reach their minimum values at the same time. Mathematically, we can also demonstrate that the rank of the coefficient matrix of the equations, including (8), (13) and the equation groups (14) with $p = 1, 2, ..., p_{\max}$, is $m$, so there is only one solution that would satisfy all these equations if it does exist.

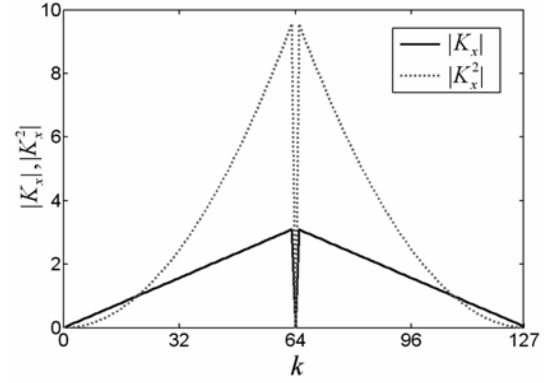

Fig.1 Magnitude of the spatial spectrums of $K_x$ and $K_x^2$ with $N = 128$.

In the following, we will show that the source pattern $[a_0, a_1, ... a_{m-1}]$ given by

$$a_l = \frac{1}{2^{m-1}} \frac{(m-1)!}{l!(m-1-l)!} \quad (17)$$

for $l = 0, 1, ..., m-1$, being the normalized $m$ th row of Pascal's triangle, is exactly the solution of (8), (13) and (14) with $p = 1, 2, ..., p_{\max}$. Firstly, according to the properties of Pascal's triangle,

$$\frac{1}{2^{m-1}}(1-1)^{m-1} = \sum_{l=0}^{m-1} a_l[(1)^{m-1-l}(-1)^l] = 0,$$

$$\frac{1}{2^{m-1}}(1+1)^{m-1} = \sum_{l=0}^{m-1} a_l[(1)^{m-1-l}(-1)^l] = 1,$$

so the two equations (8) and (13) are fulfilled. Further with (17), after doing some calculation, we find (15) becomes

$$L_{N/2-p}(a_0, ..., a_{m-1}, \lambda_p) = \cos^{m-1}[\frac{\pi}{N}(\frac{N}{2} - p)]. \quad (18)$$

Because $L_{N/2-p}$ is a constant for each $p$ under the source pattern $[a_0, a_1, ... a_{m-1}]$ determined by (17), the equation group (14) always holds true. Therefore (17) is the only solution of the optimal source patterns that can simultaneously minimize the source's high spatial frequency components, and subsequently result in much lower levels of aliasing errors introduced by the superimposed source item $S^{n+1/2}$.

### III. NUMERICAL VERIFICATION

In Fig.2, we show the spatial spectrums of some three-cell sources under different normalized patterns, the optimal pattern $[1/4, 1/2, 1/4]$, one non-optimal symmetric pattern $[0.23, 0.54, 0.23]$, the identical three-cell pattern $[1/3, 1/3, 1/3]$ and one asymmetric pattern $[2/8, 5/8, 1/8]$, for $N = 128$. It is evident that the spatial spectrum of the optimal pattern has much lower high-frequency components than the other three. Further, although being derived from the 1-D problem, the proposed optimal source patterns can be easily extended to 2-D or 3-D cases because the differencing process of each field component in PSTD is always actuated referring to only one of the three Cartesian coordinates. For example, the optimal spatially smoothed sources comprising $3 \times 3$ and $3 \times 3 \times 3$ cells, would have the normalized source patterns illustrated in Fig.3. Note that the cells in each column and row retain the proposed optimal three-cell pattern. To make our proposition more convincing, the 1-D, 2-D and 3-D practical PSTD simulations are conducted to verify the excellent performance of these optimal source patterns. For the simulations under test, the temporal driving function $f(t)$ is assumed to be a Gaussian derivative pulse with

$$f^{n+1/2} = \sin(0.1\pi n)e^{-0.002(n-100)^2},$$

and assigned to the electric field $E_z$. The sources are placed at the centers of the 1-D, 2-D and 3-D grid spaces with $N = 128$ in each coordinate and the artificial detectors are set at the specific cells $i_d = 120$, $(i_d, j_d) = (120, 64)$, and $(i_d, j_d, k_d) = (120, 64, 64)$ for 1-D, 2-D and 3-D spaces, respectively, to record the aliasing errors for the first 200 time steps before the perceptible source waves pass the detectors in theory. Fig.4 shows the simulation results for the aliasing errors of electric field from the normalized soft and hard sources with the optimal pattern $[1/4, 1/2, 1/4]$ and the non-optimal pattern $[0.249, 0.502, 0.249]$ applied for cells in each column and row. Evidently, we find that the magnitude of aliasing errors on the electric field $E_z$ introduced by the optimal sources are approximately four orders smaller than those by the non-optimal sources, which holds true for all the 1-D, 2-D and 3-D PSTD problems.

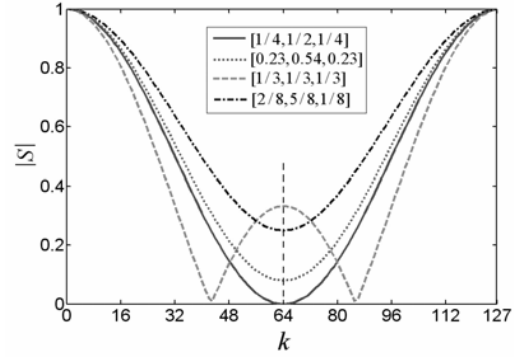

Fig.2 The spatial-spectrum magnitudes of several three-cell sources under different normalized source patterns for $N = 128$.

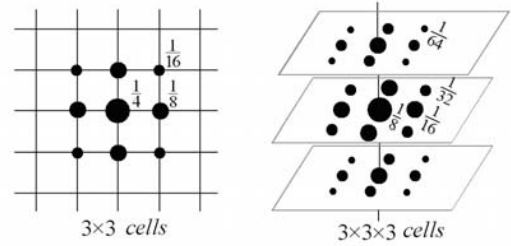

Fig.3 Normalized optimal patterns for the 2-D and 3-D sources with three cells in each column and row. The sizes of the solid dots denote their magnitudes as specified by the values nearby in the top right.

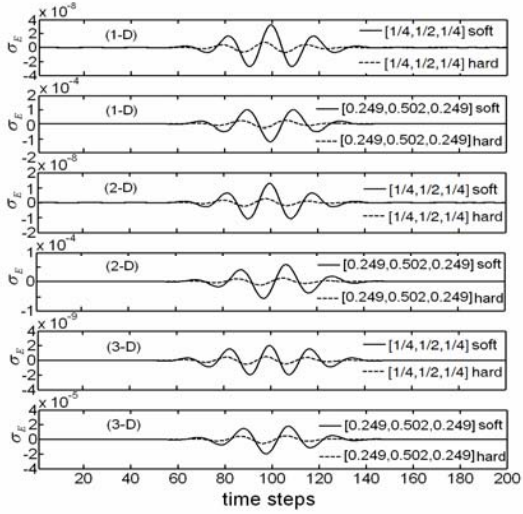

Fig.4 The introduced aliasing errors on the electric field $E_z$ by the soft and hard sources under the optimal and non-optimal patterns are shown for the first 200 time steps in the practical 1-D, 2-D and 3-D PSTD transient simulations with $c_x = 1/4$.

## IV. CONCLUSIONS

The optimal patterns of the spatially-smoothed sources are an important issue for PSTD algorithms in order to suppress the associated aliasing errors to minimum levels. In this work, we propose that the optimal pattern of a soft or hard source composed $m$ cells is exactly corresponding to the normalized $m$ th row of the Pascal's triangle. These optimal patterns deduced from 1D analysis can be easily extended to their 2D and 3D counterparts. Their excellent performance is also verified by the practical 1-D, 2-D and 3-D PSTD simulations as compared to that of non-optimal ones.